\documentclass[11pt]{article}
\usepackage[dvipdfmx]{graphicx}
\usepackage{color}
\usepackage{amssymb}
\usepackage{amsmath}
\usepackage{amsfonts}
\usepackage{algorithmic}
\usepackage{txfonts}
\usepackage{eclbkbox}
\usepackage{lscape}
\usepackage{ulem}
\usepackage{url}
\newtheorem{definition}{Definition}

\newcommand{\defeq}{\overset{\mathrm{def}}{=}}

\newcommand*\samethanks[1][\value{footnote}]{\footnotemark[#1]} 

\title{On Practical Accuracy of Edit Distance Approximation Algorithms}
\author{
	Hiroyuki Hanada
	\thanks{Graduate School of Information Science and Technology, Hokkaido University.
	Kita 14, Nishi 9, Kita-ku, Sapporo, Hokkaido, 060-0814, Japan.
	(The affiliations of all authors at which the research is conducted,
	 and current affiliation of Mineichi Kudo and Atsuyoshi Nakamura)}
	\thanks{Department of Computer Science, Nagoya Institute of Technology.
	Gokiso-cho, Showa-ku, Nagoya, Aichi, Japan.
	(Current affiliation of Hiroyuki Hanada)}
	\\
	{\tt hana-hiro@live.jp}
	\and
	Mineichi Kudo
	\samethanks[1]
	\\
	{\tt mine@main.ist.hokudai.ac.jp}
	\and Atsuyoshi Nakamura
	\samethanks[1]
	\\
	{\tt atsu@main.ist.hokudai.ac.jp}
}
\date{}

\begin{document}
\maketitle
\normalem

\begin{abstract}
The edit distance is a basic string similarity measure used in many applications such as text mining, signal processing, bioinformatics, and so on. However, the computational cost can be a problem when we repeat many distance calculations as seen in real-life searching situations. \\
A promising solution to cope with the problem is to approximate the edit distance by another distance with a lower computational cost. There are, indeed, many distances have been proposed for approximating the edit distance. However, their approximation accuracies are evaluated only theoretically: many of them are evaluated only with big-oh (asymptotic) notations, and without experimental analysis. Therefore, it is beneficial to know their actual performance in real applications.\\
In this study we compared existing six approximation distances in two approaches: (i) we refined their theoretical approximation accuracy by calculating up to the constant coefficients, and (ii) we conducted some experiments, in one artificial and two real-life data sets, to reveal under which situations they perform best. As a result we obtained the following results: [Batu 2006] is the best theoretically and [Andoni 2010] experimentally. Theoretical considerations show that [Batu 2006] is the best if the string length $n$ is large enough ($n\geq 300$). [Andoni 2010] is experimentally the best for most data sets and theoretically the second best. [Bar-Yossef 2004], [Charikar 2006] and [Sokolov 2007], despite their middle-level theoretical performance, are experimentally as good as [Andoni 2010] for pairs of strings with large alphabet size.

\end{abstract}

\section{Introduction} \label{ch:overview}
%

The edit distance between two strings $x$ and $y$, denoted by $d_e(x, y)$ in this paper, is defined by the minimum number of character-wise edit operations (insertions, deletions or substitutions) to identify $x$ and $y$ (Section \ref{ch:defs-strings}). The distance has been intensively researched because it naturally fits for many real-life situations: error detection in documents, noise analysis in signal processing, mutation-tolerant database searching in genomes and proteins, and so on \cite{suftreebook,GuideToApproxMatching}.

A weak point of the edit distance is its quadratic computation cost $O(n^2)$, where $n$ is the string length to be compared. Many efforts, therefore, have been devoted to reduce the cost. They are separated by whether approximations of the distance are conducted or not. Unless some approximation is made, it is hard to reduce the worst-case computational cost from $O(n^2)$. Some methods without approximation \cite{ParallelSerialApproxMatch,SublinearAlgoApproxSearch} achieve the worst-case computational time $O(nk)$, where $k$ is the maximum edit distance to be considered. This means $O(n)$ if $k$ is a constant; but $O(n^2)$ in the worst case because $k$ can be $n$. Only approximation methods can achieve a linear or quasi-linear time such as $O(n^{1+\varepsilon})$ or $O(n(\log n)^m)$. Then the next question with some approximation algorithms is whether they have sufficiently good approximation accuracy or not.

To answer the question, we will do in this paper the following studies:
\begin{description}
\item[Theoretical evaluations:] We consider the {\it distortion} (Section \ref{ch:defs-approx}) as a typical measure of approximation accuracy. Many approximation algorithms (four out of six in this paper) conducted only big-oh (asymptotic) analyses in the distortion, for example, $O(n\log n)$ rather than $100 n\log n$. However, in real-life situations, non-asymptotic distortions are desired. So we refine the analyses so as to reveal the constant factors.
\item[Experimental evaluations:] Most existing methods (all of six in this paper) have not received any experimental evaluation on the approximation accuracy. So we examine their experimental accuracy in three datasets (one artificial and two real).
\end{description}

\section{Preparation}

\subsection{Definitions for strings} \label{ch:defs-strings}

Throughout the paper, by $\Sigma$ we denote the alphabet (the set of characters). Let $\Sigma^n$ be the set of all strings of length $n$.

For a string $x$, we denote by $|x|$ the length of $x$, by $x[i]$ the $i$th character of $x$, and by $x[i..j]$ the substring of $x$ consisting of its $i$th to $j$th characters. A {\em $q$-gram} is a substring of length $q$.

The {\it edit distance} \cite{suftreebook} $d_e(x, y)$ for two strings $x, y$ is defined by the minimum number of edit operations: inserting, deleting or substituting one character in $x$ to make $x$ be identical to $y$.

\subsection{Distortion}

\subsubsection{Definition} \label{ch:defs-approx}

We use the {\em distortion}, also known as the {\em approximation factor}, as a measure of approximation accuracy of a function defined as follows:

\begin{definition} \cite{LectureDiscreteGeometry}\cite{EmbedUlamMetric} \label{df:def-distortion}
Given a set $S$, a non-negative function $f(z)$ and a non-negative approximation function $\tilde{f}(z)$, the {\em distortion} of $\tilde{f}(z)$ to $f(z)$ is defined by the smallest $K\in[1,+\infty)$ such that
\[
\exists K'\in(0,+\infty), \forall z\in S: f(z) \leq K' \tilde{f}(z) \leq K f(z).
\]
\end{definition}

\begin{figure}[t]
\begin{center}
\includegraphics[width=9cm,clip]{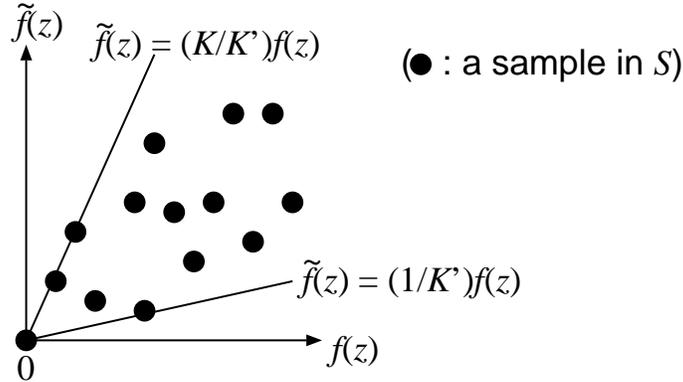}
\vspace{-1em}
\caption{The concept of distortion $K$ over a set $S$}
\label{fg:distortion-figure}
\vspace{-1em}
\end{center}
\end{figure}

The concept is illustrated in Fig. \ref{fg:distortion-figure}. Note that, in this paper, $S$ is given as a set of pairs of strings $\{z = (x, y)\}$ since we consider $f(z) = d_e(x, y)$ and $\tilde{f}(z) = \tilde{d_e}(x, y)$, where $\tilde{d_e}(x, y)$ is a string distance approximating $d_e(x, y)$. The value of $K$ shows the ratio of the upper bound $(K/K')f(z)$ to the lower bound $(1/K')f(z)$. A smaller value of distortion $K$ $(\geq 1)$, therefore, means better approximation. Especially, $K=1$ means that $f(z)$ and $\tilde{f}(z)$ are proportional to each other.

%

\subsubsection{Asymptotic/non-asymptotic distortion analysis} \label{ch:analyses-approx}

\begin{figure}[tp]
\begin{center}
\begin{tabular}{ccc}
\begin{minipage}[t]{4.3cm}
\begin{center}
\includegraphics[width=4cm,clip]{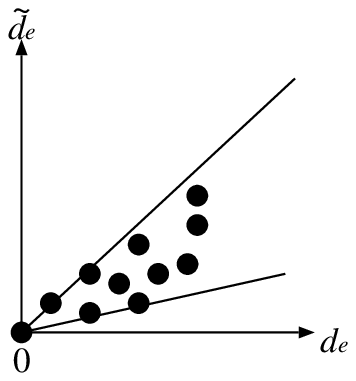}

(a) $n$ is small
\end{center}
\end{minipage}
&
\begin{minipage}[t]{4.3cm}
\begin{center}
\includegraphics[width=4cm,clip]{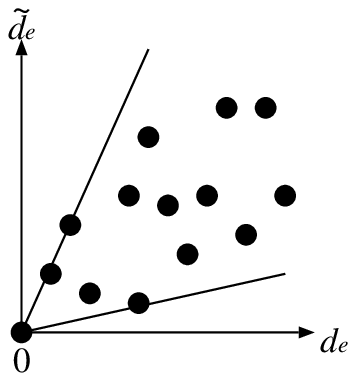}

(b) $n$ is large
\end{center}
\end{minipage}
&
\begin{minipage}[t]{4.3cm}
\begin{center}
\includegraphics[width=4cm,clip]{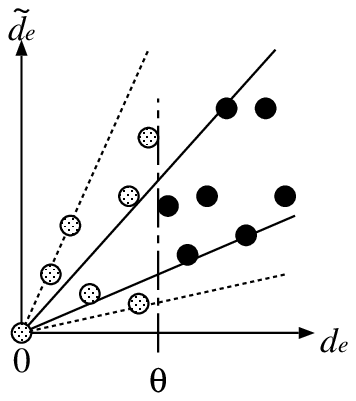}

(c) $d_e(x, y)\geq\theta$
\end{center}
\end{minipage}
\end{tabular}
\caption{Several situations in distortion evaluation}
\label{fg:distortion-by-length}
\vspace{-1em}
\end{center}
\end{figure}

The distortion is an intuitive measure for showing how close the value of the approximation distance $\tilde{d_e}(x, y)$ is to the original distance $d_e(x, y)$. However, we have to pay attention to what the distortion actually means in several conditions (Fig. \ref{fg:distortion-by-length}).

First we notice that the value of distortion, in general, becomes larger as the string length $n$ increases, assuming $|x| = |y| = n$ (Fig. \ref{fg:distortion-by-length}, (a) and (b)).
Taking this tendency into account, many of existing papers evaluate the distortions by big-oh notations, that is, how slowly the value $K$ increases as $n$ increases.

We should also notice another tendency that the distortion is often affected strongly by string pairs with a small value of $d_e$ (Fig. \ref{fg:distortion-by-length}(b)(c)). To ignore such an exceptional situation, some of the existing methods are evaluated only in the range of $d_e\geq\theta$ with a threshold $\theta$ (Fig. \ref{fg:distortion-by-length}(c)).
%
%
%
%

\section{Outline of existing approximation methods} \label{ch:methods}

We chose six approximation algorithms to be compared from the two viewpoints: coverage of almost all state-of-the-art algorithms and implementation easiness. We explain those algorithms in four groups according to their characteristics.

\begin{description}
\item[$q$-gram-based algorithms] (two of: \cite{ApproxEditDistEfficiently}=[Bar-Yossef 2004], \cite{VectorRepSimilarString}=[Sokolov 2007]) \\
	These two algorithms approximate the edit distance by counting occurrences of $q$-grams in given two strings, and then take the difference between them.
\item[Ulam-metric-based algorithms] (two of: \cite{EmbedUlamMetric}=[Charikar 2006], \cite{L1NonEmbeddability}=[Andoni 2009]) \\
	These two algorithms are originally developed for the {\it Ulam metric}, which is the edit distance in the set of strings whose characters are all distinct \cite{EmbedUlamMetric}. It can be shown that the Ulam metric is applicable for the edit distance between general strings with some simple operations (Section \ref{ch:experiment-procedure}). The distance computation of the two algorithms exploits the property that every string does not contain the same character twice or more. For example, in [Charikar 2006], the distance is defined as the sum of $|1/(x^{-1}[b] - x^{-1}[a]) - 1/(y^{-1}[b] - y^{-1}[a])|$ for all pairs $(a, b)\in\Sigma\times\Sigma$, where $x^{-1}[a]$ denotes the position of $a$ found in the string $x$ (omitted if $a$ is not in $x$).
\item[Restricted alignment algorithms] (one of: \cite{PolylogApproxEditDist-full}=[Andoni 2010]) \\
	The edit distance can be regarded as a character-wise {\it alignment} between two strings \cite{suftreebook}. [Andoni 2010] uses $q$-gram-wise alignment instead and assures certain approximation accuracy even if a pruning in the calculation is conducted\footnote{The algorithm of [Andoni 2010] needs $O(n^2)$ time if no pruning is made, which is equal to that of the edit distance.}.
\item[Shrinking algorithms] (one of: \cite{ObliviousStrEmbed}=[Batu 2006]) \\
	Batu's algorithm converts given strings into shorter ones by merging some characters into one such as ``abcbbabc'' $\to$ ``XYX'' with the rule ``abc'' $\to$ ``X'' and ``bb'' $\to$ ``Y''. Then it computes the edit distance of the converted strings as the approximated distance.
\end{description}

\section{Refined theoretical distortions} \label{ch:theoretical-distortions}

\subsection{Outline}

\begin{table}[tp]
\rotatebox{90}{
\begin{minipage}{\textheight}
\begin{center}
\caption{Refined distortions. Here, $\tilde{d_e}$ is the approximated distance of $d_e$; the strings are limited to length $n$; a threshold $\theta$ is employed to limit $d_e\geq\theta$ in some algorithms. In logarithms, the bases are $2$ for $\lg$ and $e$ for $\ln$, respectively.}
\label{tb:existing-methods-orig}
\begin{tabular}{cccc}
\hline
Algorithm & Original distortion & Original inequality & Refined distortion \\
\hline
{}[Bar-Yossef 2004] \cite{ApproxEditDistEfficiently}
	&&
\begin{minipage}[c]{0.23\hsize}
\vspace{-1em}
\begin{eqnarray*}
\begin{cases}
d_e\leq k \Rightarrow \tilde{d_e}\leq 4kq,{}^{\dagger} \\
d_e\geq 13(kn)^{\frac{2}{3}} \Rightarrow \tilde{d_e}\geq 8kq
\end{cases}
\end{eqnarray*}
\end{minipage}
	& $\frac{13}{2\theta^{1/3}}n^{2/3}$ \\
{}[Batu 2006] \cite{ObliviousStrEmbed}
	& $\min\left\{n^{\frac{1}{3}+o(1)}, (d_e)^{\frac{1}{2} + o(1)}\right\}$
	&& $4(2c-1)\left(\lg((2c-3)k) + 1 + \frac{(c - 1)^2}{c}\right){}^{\ddagger}$ \\
{}[Charikar 2006] \cite{EmbedUlamMetric}
	& $O(n\log n)${}${}^{\dagger\dagger}$ && $48n(1+\ln n)/\max\{1,\theta\}$ \\
{}[Sokolov 2007] \cite{VectorRepSimilarString}
	&&
\begin{minipage}[c]{0.23\hsize}
\vspace{-1em}
\begin{eqnarray*}
\begin{cases}
d_e\leq k \Rightarrow \tilde{d_e}\leq \frac{2k(n+2)}{n}, \\
d_e > k \Rightarrow \tilde{d_e}\geq \frac{2k - 8}{n}
\end{cases}
\end{eqnarray*}
\end{minipage}
	&
\begin{minipage}[c]{0.23\hsize}
\vspace{-1em}
\begin{eqnarray*}
\begin{cases}
+\infty & (\theta\leq 5),\\
\frac{n\theta + 2}{\theta - 5} & (\theta > 5)
\end{cases}
\end{eqnarray*}
\end{minipage}
	\\
{}[Andoni 2009] \cite{L1NonEmbeddability}
	& $O(n)${}${}^{\dagger\dagger}$ && $3400n$ \\
{}[Andoni 2010] \cite{PolylogApproxEditDist-full}
	& $12\lg n{}^{\ddagger\ddagger}$ && $12\lg n{}^{\ddagger\ddagger}$ \\
\hline
\end{tabular}
\end{center}
{\footnotesize
Note:
\begin{itemize}
\item[$\dagger$] $q$ denotes the $q$-gram. In the algorithm, $q$ is set to $n^{2/3}/(2k^{1/3})$.
\item[$\ddagger$] $c = \max\{(\lg\lg n)/(\lg\lg\lg n), 2\}$.
\item[$\dagger\dagger$] In [Charikar 2006] and [Andoni 2009], the distortions are derived for the Ulam metric as $O(\log n)$ and $O(1)$, respectively. We multiplied them by $O(n)$ (more precisely, $2n$) so as to be applicable to general strings (Section \ref{ch:experiment-procedure}).
\item[$\ddagger\ddagger$] The distortion is shown in the original paper (\cite{PolylogApproxEditDist-full}, pp. 16 in the full version).
\end{itemize}
}
\end{minipage}
}
\end{table}

We re-analyzed the six algorithms to obtain their distortions with constant factors. The results are shown in Table \ref{tb:existing-methods-orig}.

Before analyzing the table in detail in Section \ref{ch:compare-theoretical-distortions}, in Section \ref{ch:calc-distortion-asympt} we explain how the constant factors are extracted from big-oh notations, and how the accuracy evaluations with inequalities are converted to distortions with a threshold $\theta$.

\subsection{Derivation of distortions} \label{ch:calc-distortion-asympt}

For each algorithm whose distortion is given in a big-oh notation ([Batu 2006], [Charikar 2006], [Andoni 2009] and [Andoni 2010]), we examined every step in the algorithm. The detailed derivations are given in Appendix \ref{ch:distortion-re-analyses}.
%
%

\begin{figure}[tp]
\begin{center}
\begin{tabular}{cc}
%
%
\begin{minipage}[t]{5.8cm}
\begin{center}
\includegraphics[width=5.5cm,clip]{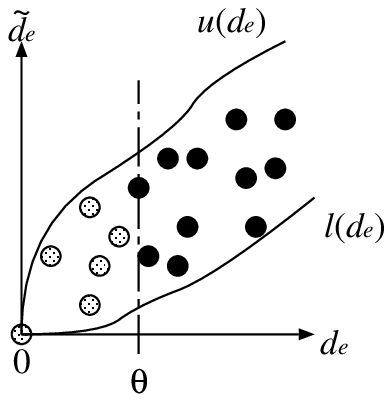}

(a) $l(d_e)\leq\tilde{d_e}\leq u(d_e)$
\end{center}
\end{minipage}
&
\begin{minipage}[t]{6.5cm}
\begin{center}
\includegraphics[width=6.2cm,clip]{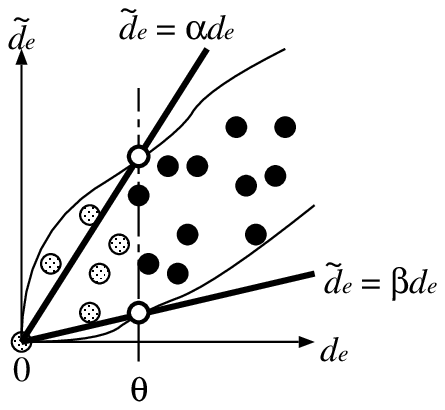}

(b) Distortion for $d_e\geq\theta$
\end{center}
\end{minipage}
\end{tabular}
\caption{Conversion of lower and upper bounds to a distortion}
\label{fg:ineq2distortion-curve}
\vspace{-1em}
\end{center}
\end{figure}

For each algorithm whose accuracy is bounded by inequalities ([Bar-Yossef 2004] and [Sokolov 2007]), we calculated its distortion by the following procedure. Detailed distortion calculations for the two algorithms are shown in Appendix \ref{ch:distortion-from-ineq}.

Let $\tilde{d_e}$ be bounded by two functions of $d_e$ as $l(d_e)\leq\tilde{d_e}\leq u(d_e)$ for $d_e\geq\theta$ (Fig. \ref{fg:ineq2distortion-curve}(a)). Then the distortion $K$ of $\tilde{d_e}$ for $d_e\geq\theta$ is upper-bounded by $K_{\theta} = u(\theta)/l(\theta)$ under the monotonicity of slopes $u(d_e)/d_e$ and $l(d_e)/d_e$. Indeed, if $u(d_e)/d_e$ and $l(d_e)/d_e$ are monotonically decreasing and increasing in $d_e\geq\theta$, respectively, then $K = (\sup_{d_e\geq\theta}u(d_e)/d_e)/(\inf_{d_e\geq\theta}l(d_e)/d_e)\leq (u(\theta)/\theta)/(l(\theta)/\theta) = u(\theta)/l(\theta) = K_{\theta}$. Therefore we can obtain the distortion when the monotonicity of them are confirmed.
\subsection{Comparison of calculated distortions} \label{ch:compare-theoretical-distortions}

Now we examine the refined distortions shown in Table \ref{tb:existing-methods-orig}. We note that all these algorithms can be now compared in a unified expression.

First we classify these algorithms in the complexity order. Note that we can assume that $\theta$ takes an order between $O(1)$ and $O(n)$ since the edit distance takes a value between $0$ and $n$. Assuming $\theta = O(1)$ as an ordinary case, they are ordered as:
\begin{itemize}
\item Sub-logarithmic ($O((\log\log n)^2)$): [Batu 2006]
\item Logarithmic ($O(\log n)$): [Andoni 2010]
\item Sublinear ($O(n^{\alpha(<1)})$): [Bar-Yossef 2004]
\item Linear ($O(n)$): [Sokolov 2007], [Andoni 2009]
\item Super-linear ($O(n\log n)$): [Charikar 2006]
\end{itemize}
Therefore, [Batu 2006] is the best for $\theta = O(1)$ then [Andoni 2010] follows. For $\theta = O(n)$, [Charikar 2006] also has the same logarithmic order. Thus [Charikar 2006] and [Andoni 2010] are comparable for $\theta = O(n)$.

\begin{figure}[tp]
\begin{center}
\includegraphics[width=11.5cm,clip]{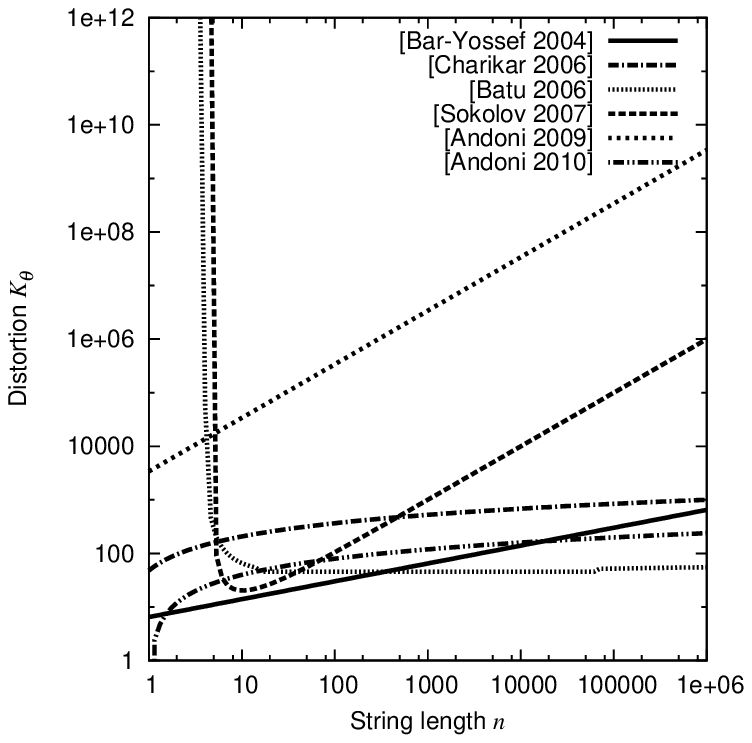}
\caption{Distortions of six approximation methods.}
\label{fg:distortion-result}
{\footnotesize
Note: $\theta = n$ for [Bar-Yossef 2004], [Charikar 2006] and [Sokolov 2007].
}
\end{center}
\end{figure}

Next let us compare the distortions in more detail. Since the refined distortions reveal the constants, we can compare algorithms for every specific value of $n$. We show the result in Fig. \ref{fg:distortion-result}. In the figure we set $\theta = n$ (maximum $\theta$) for [Bar-Yossef 2004], [Charikar 2006] and [Sokolov 2007] to evaluate optimistic distortion values.
It is observed as expected that [Batu 2006] outperforms the others if $n$ is large enough. However, when $n$ is not so large, say, $n\leq 300$, [Bar-Yossef 2004] is the best. Such a range of effective $n$ is not obtained until our analyses made clear the constant factors.

Focusing on the absolute value of distortion, it ranges from 10 to 100 for $100\leq n\leq 10000$. We might need to investigate whether such large values are acceptable in real-life applications, keeping in mind that they are evaluated in the worst case.

\definecolor{gray7}{gray}{0.467} 
\definecolor{grayA}{gray}{0.667} 

%
%
%
\section{Experimental comparison} \label{ch:experimental-distortions}

\subsection{Procedure} \label{ch:experiment-procedure}

Next we compared them experimentally to know their practical usefulness.

For each data set that will be explained in detail later, we make ready a set $S$ of 10,000 pairs of strings $S = \{(x_1, y_1)$, $(x_2, y_2)$, $\ldots$, $(x_{10000}, y_{10000})\}$. We computed the distortion for $S$ for the six approximation distances.

We used one artificial and two real-life data sets as follows:
\begin{description}
\item[Random] ($n\in\{100, 300, 1000\}$, $|\Sigma|\in\{4, 20\}$, $e\in\{4, 30\}$): \\
	First we choose $x$ from $\Sigma^n$ at random with equal probability and initialize $y$ by $x$. Then we modify $y$ until the total operation cost becomes $e$: (a) replace a randomly chosen character in $y$ with a randomly chosen character from $\Sigma$ (probability: 2/3, cost: 1) or (b) delete a randomly chosen character in $y$ and then insert a randomly chosen character at a randomly chosen position (probability: 1/3, cost: 2), where all random choices of characters and positions are conducted with equal probability. Note that $d_e(x, y)$ equals $e$ in most cases but can be less than $e$.
\item[DDBJ] ($n\in\{100, 300, 1000\}$): \\
	DDBJ (DNA Data Bank of Japan) is a DNA nucleobase sequence database service \cite{DDBJ}. We used ``ddbjhum1'' data ($|\Sigma|=15$; 4 of them occupy 99.95\%). To unify the string length in each data set, we constructed the data set as follows: For $n = 100$, we gathered strings of length 100 to 299 in ddbjhum1 and truncated the 101st character or after. Similarity, for $n = 300$ and $n = 1000$, we collected strings of length 300 to 999 for $n=300$ and 1000 to 2999 for $n=1000$, respectively.
\item[UniProt] ($n\in\{100, 300, 1000\}$): \\
	UniProt (Universal Protein Resource) is an amino acid sequence (i.e. protein) database service \cite{UniProt}. We used ``UniProtKB-SwissProt'' data ($|\Sigma|=25$; 20 of them occupy 99.99\%). We conducted the data set constructions in the same manner as in DDBJ.
\end{description}

For the algorithms assuming the Ulam metric ([Charikar 2006] and [Andoni 2009], Section \ref{ch:methods}), where all characters in a string are expected to be distinct, we ``expanded'' the alphabet from $\Sigma$ to $\Sigma^t$ for each string pair $x, y$ so that $(x[1..t], \dots, x[n-t+1..n])$ are distinct and so do $(y[1..t], \dots, y[n-t+1..n])$ with as small $t$ as possible. It can be shown that the distortion with this expansion is at most $2t$ times that under the Ulam metric \cite{EmbedUlamMetric}.

When algorithms have parameters ([Bar-Yossef 2004], [Batu 2006], [Sokolov 2007] and [Andoni 2010]), we chose the smallest distortions over some candidates of parameters as follows:
\begin{itemize}
\item $q\in\{2, 4, 6\}$ for $q$-grams ([Bar-Yossef 2004] and [Sokolov 2007]\footnote{Following the description in the papers [Bar-Yossef 2004] and [Sokolov 2007], $B$ and $q_1$ corresponds to $q$, respectively. In [Sokolov 2007], parameter $q_2$ is also set to $q$.}).
\item $c\in\{2, 4\}$ and $j = 1$ for [Batu 2006] (see Appendix \ref{ch:distortion-re-analyses} for details). As a result, the theoretical distortion of [Batu 2006] is $(2c-1)\cdot [4c + \{8(2c-3)k\}^{c-1}]/c = 12[1 + \lceil\lg|\Sigma|\rceil] = 72$ with $c = 2$ and $|\Sigma| = 20$, a constant against $n$. It needs $O(n^2)$ time.
\item Tree node pruning (the trade-off between the computational time and the accuracy) is not conducted on [Andoni 2010] (the highest accuracy). It needs $\Omega(n^2)$ time.
\end{itemize}

\subsection{Results} \label{ch:exp-results}

\begin{figure}[tp]
\begin{center}
\small{
\begin{tabular}{cc}
\includegraphics[width=6cm,clip]{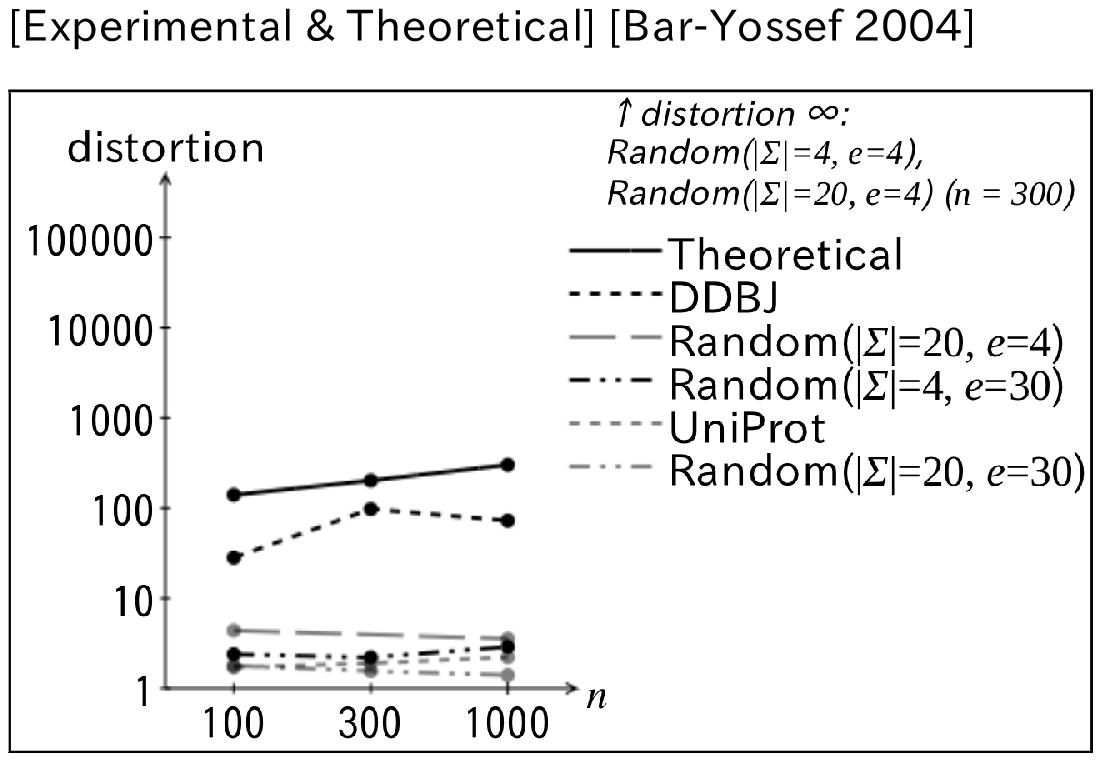}
&
\includegraphics[width=6cm,clip]{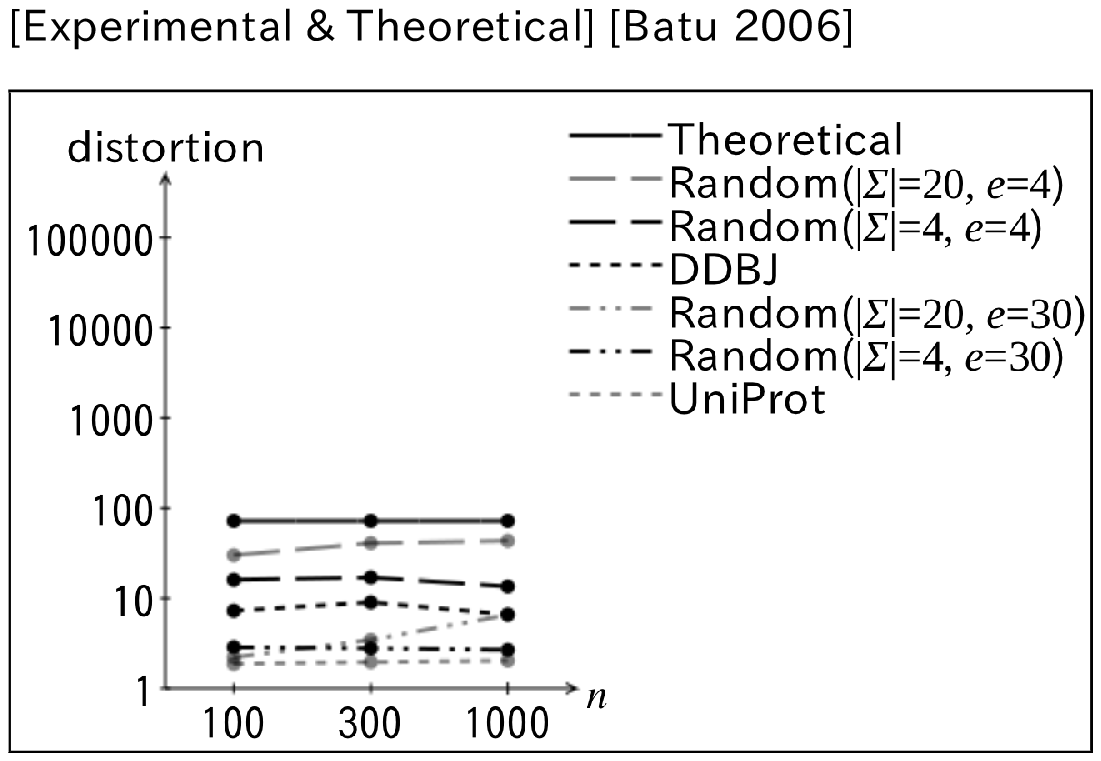}
\\
(a) [Bar-Yossef 2004] & (b) [Batu 2006]
\\
\\
\includegraphics[width=6cm,clip]{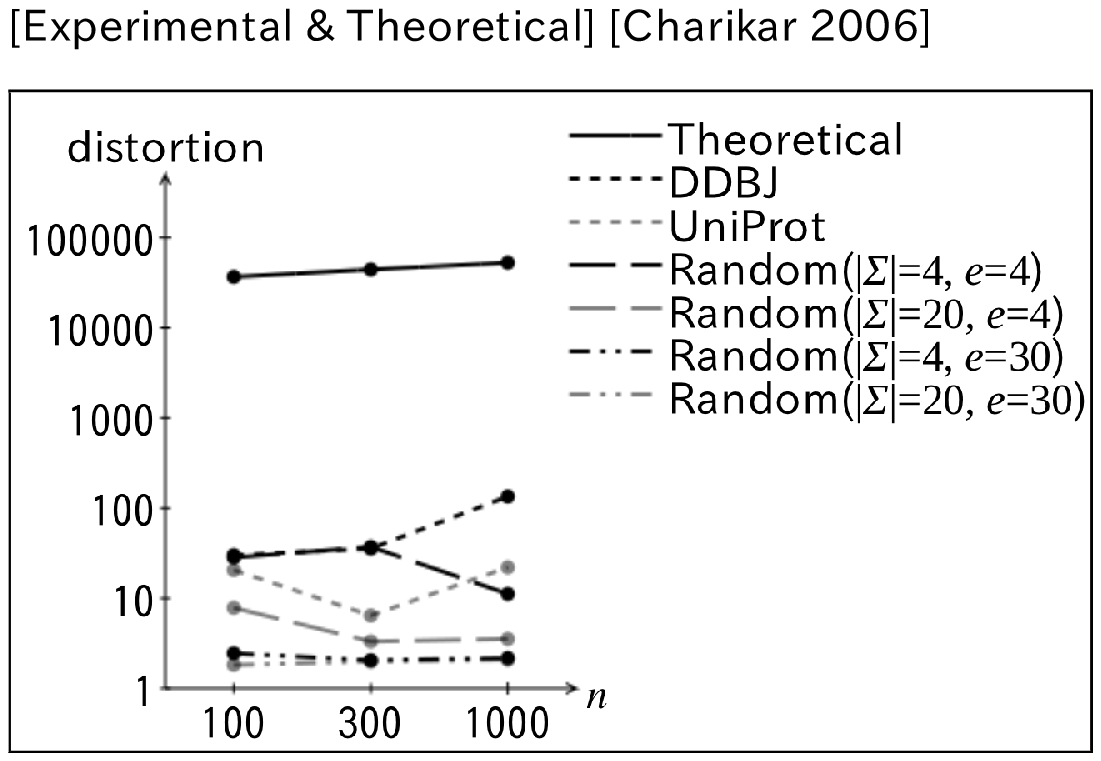}
&
\includegraphics[width=6cm,clip]{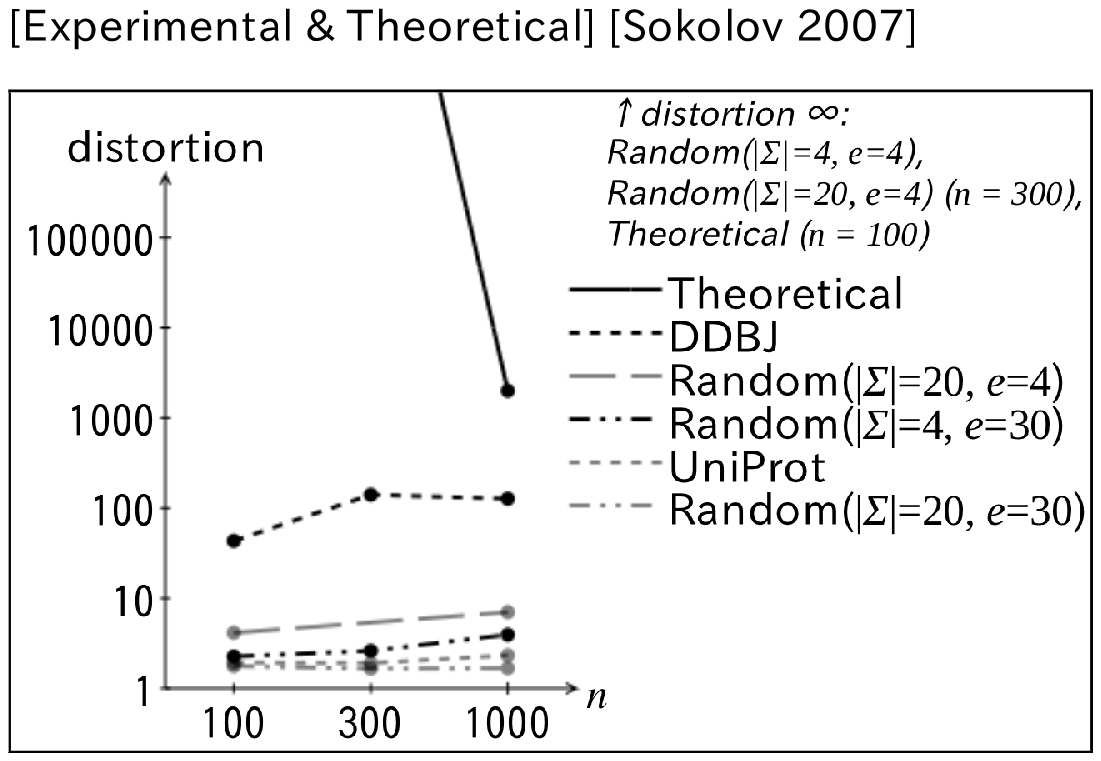}
\\
(c) [Charikar 2006] & (d) [Sokolov 2007]
\\
\\
\includegraphics[width=6cm,clip]{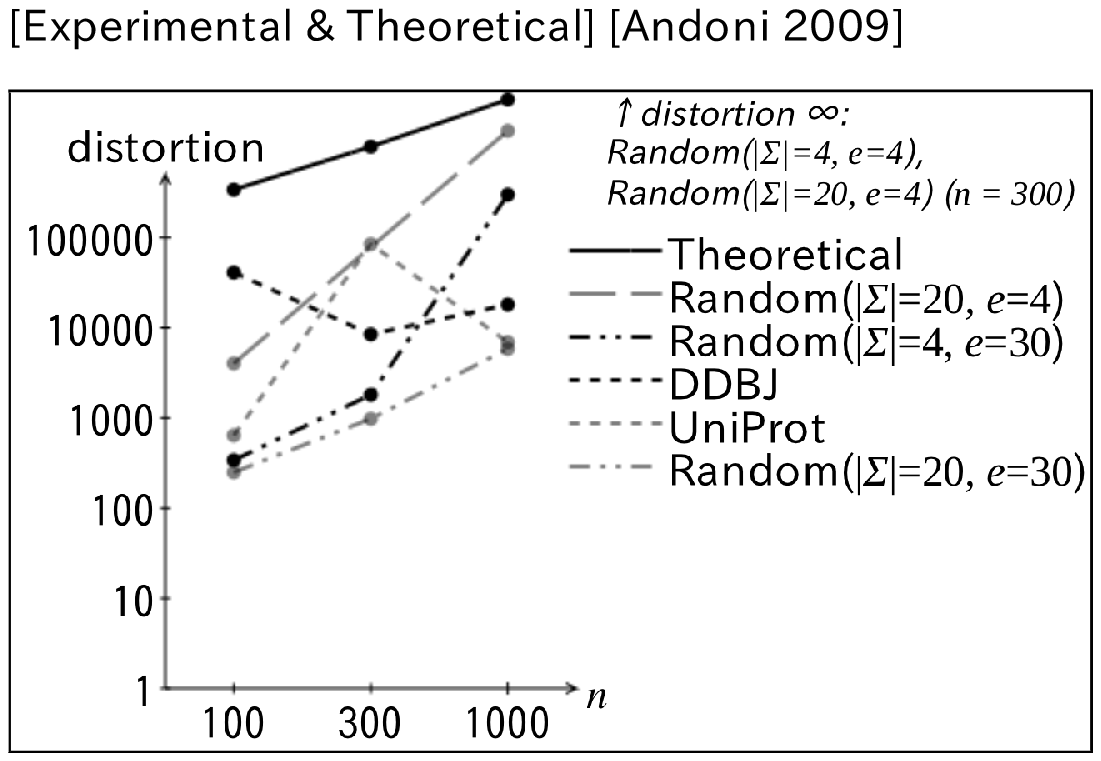}
&
\includegraphics[width=6cm,clip]{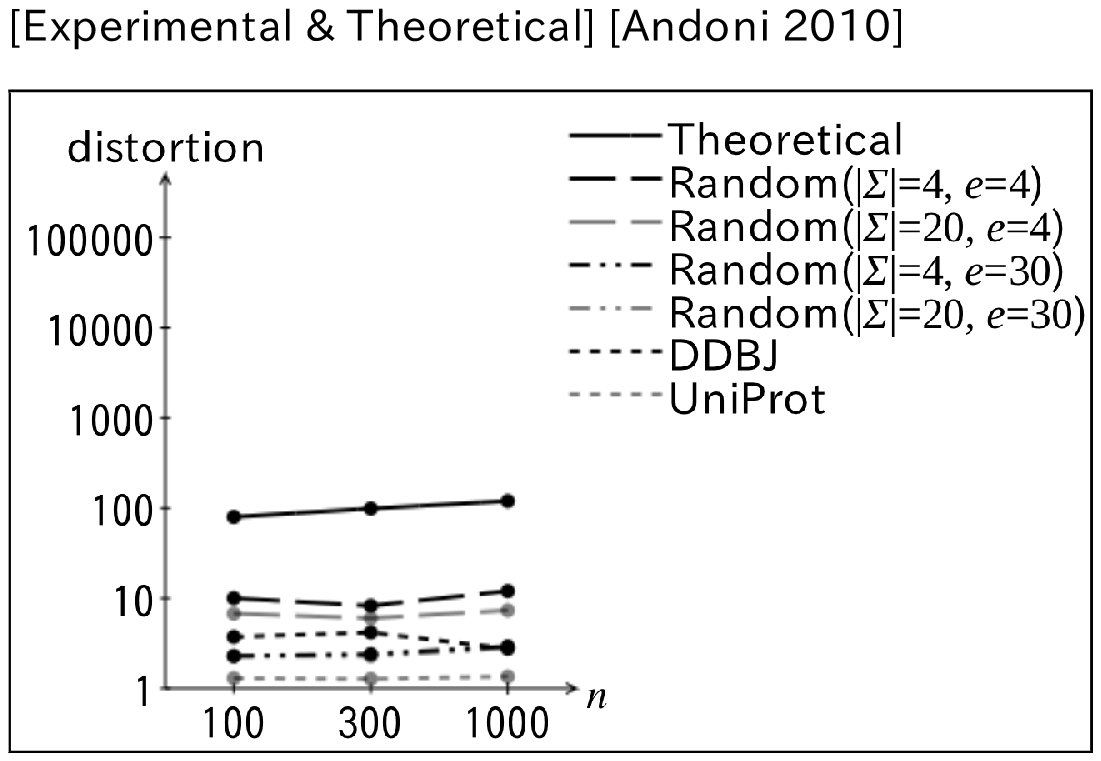}
\\
(e) [Andoni 2009] & (f) [Andoni 2010]
\end{tabular}
}
\end{center}
\caption{Experimental distortions of six algorithms. Gray lines denote $|\Sigma|=20$ data sets including UniProt. The theoretical value of [Batu 2006] is different from that in Table \ref{tb:existing-methods-orig} (constant against $n$; see Section \ref{ch:experiment-procedure}).}
\label{fg:exp-algorithms}
\end{figure}

\begin{figure}[tp]
\begin{center}
\small{
\begin{tabular}{cc}
\includegraphics[width=6cm,clip]{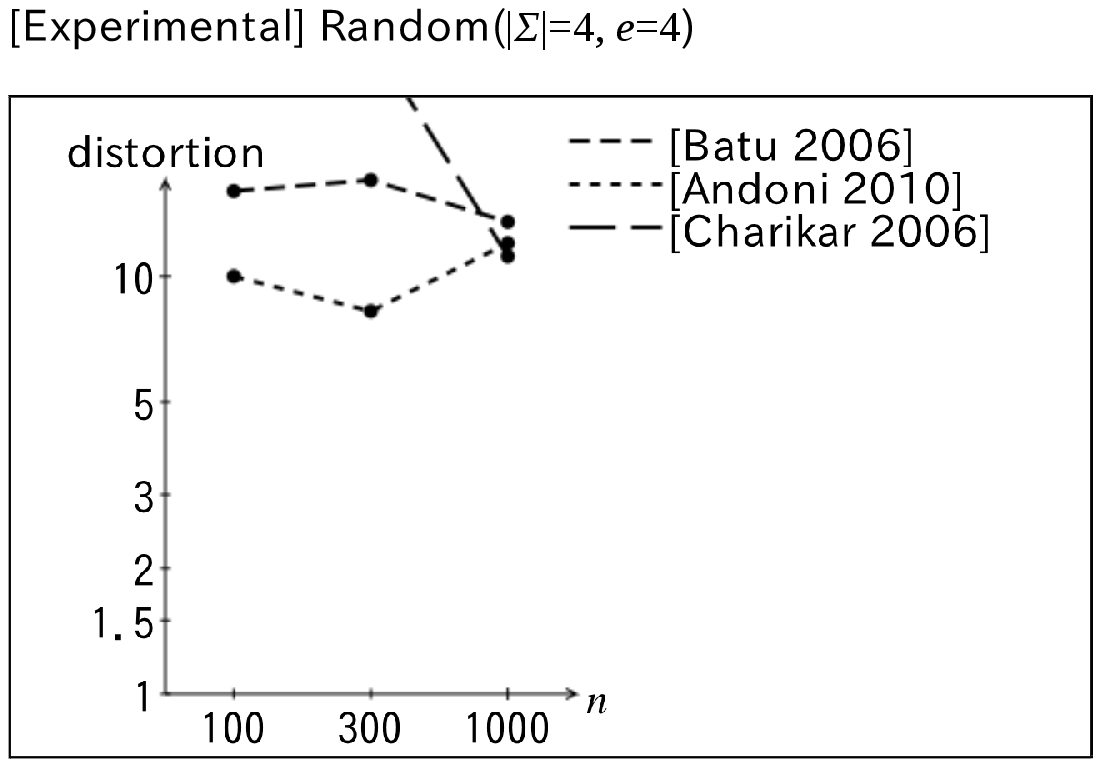}
&
\includegraphics[width=6cm,clip]{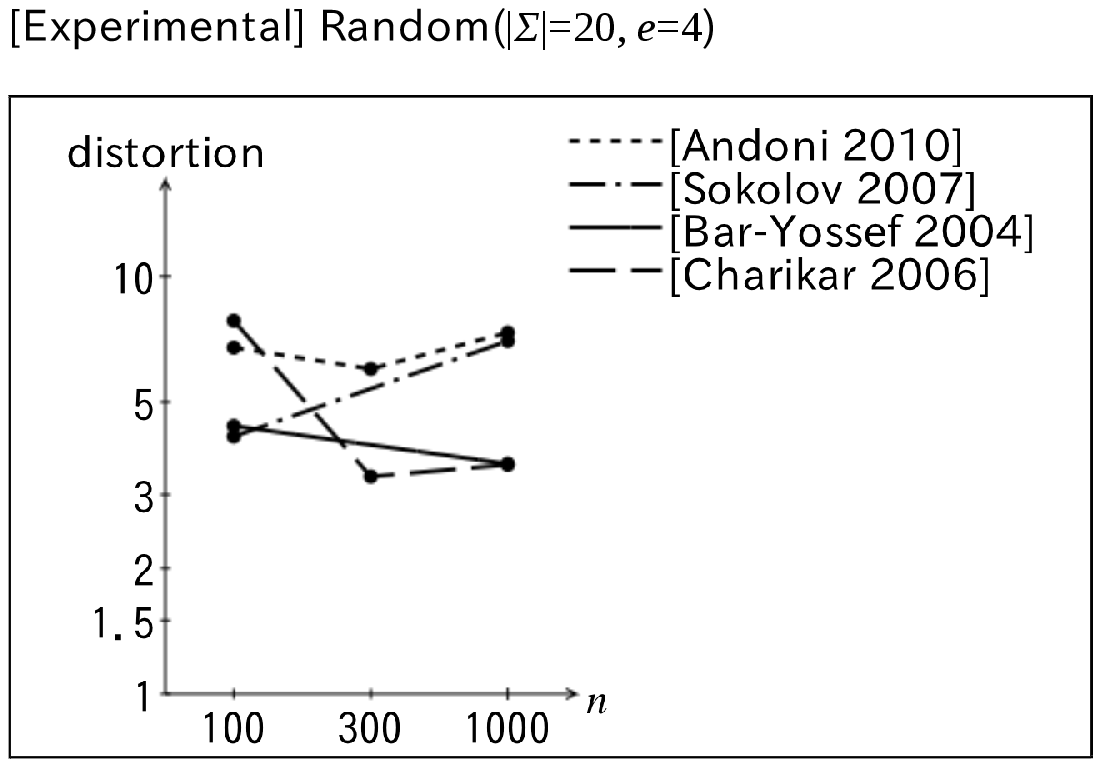}
\\
(a) Random, $|\Sigma|=4$, $e=4$ & (b) Random, $|\Sigma|=20$, $e=4$
\\
\\
\includegraphics[width=6cm,clip]{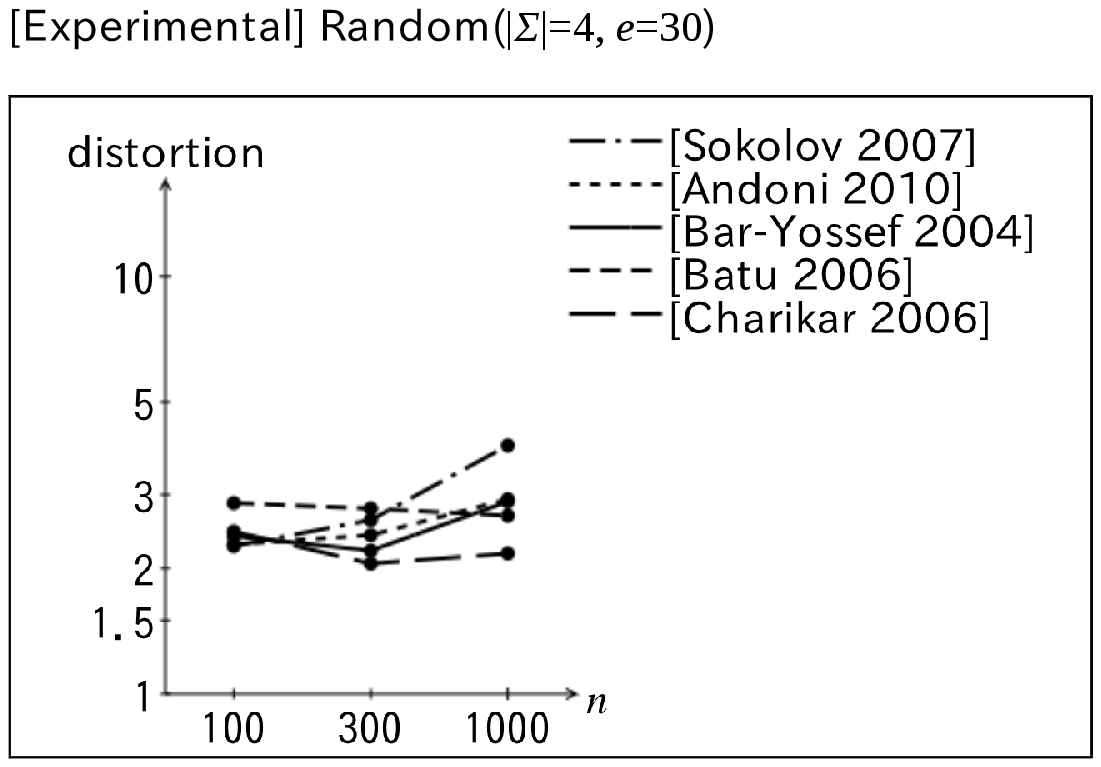}
&
\includegraphics[width=6cm,clip]{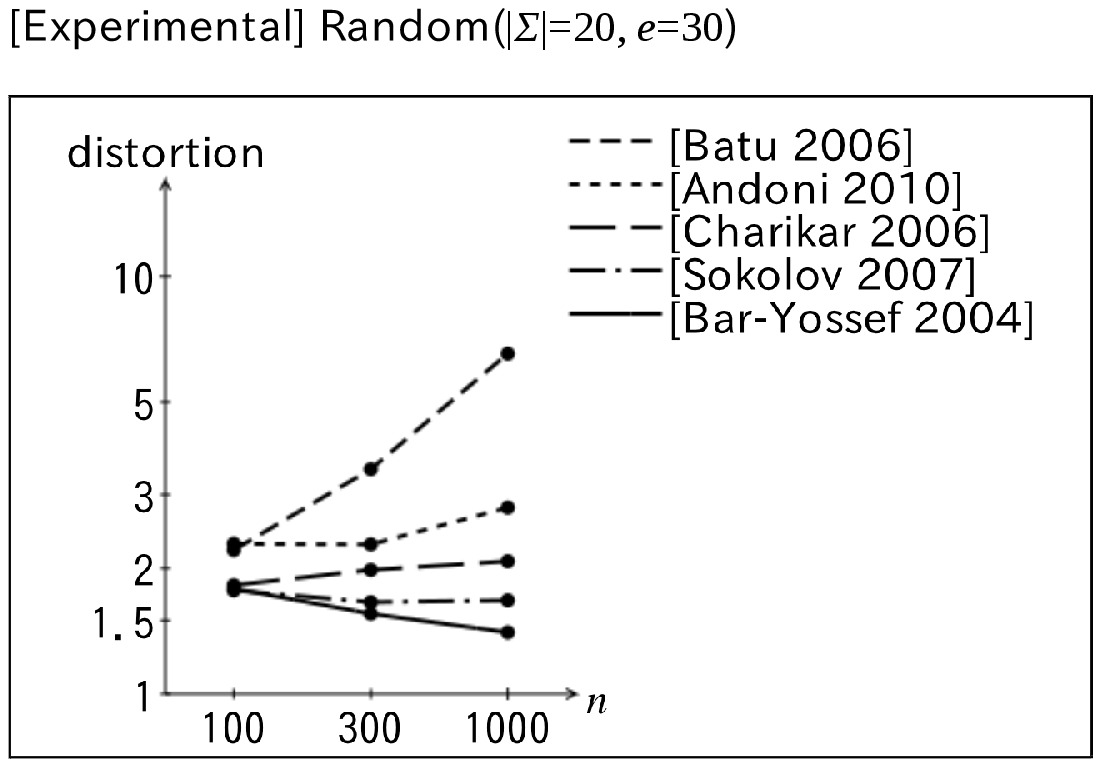}
\\
(c) Random, $|\Sigma|=4$, $e=30$ & (d) Random, $|\Sigma|=20$, $e=30$
\\
\\
\includegraphics[width=6cm,clip]{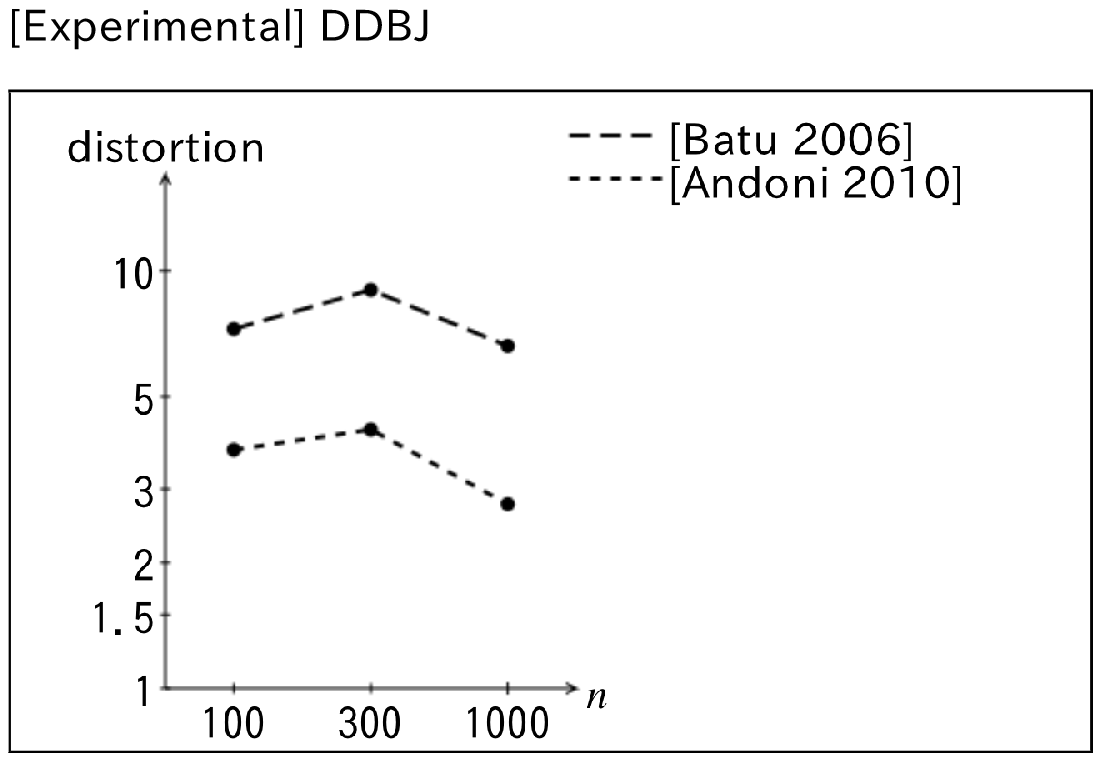}
&
\includegraphics[width=6cm,clip]{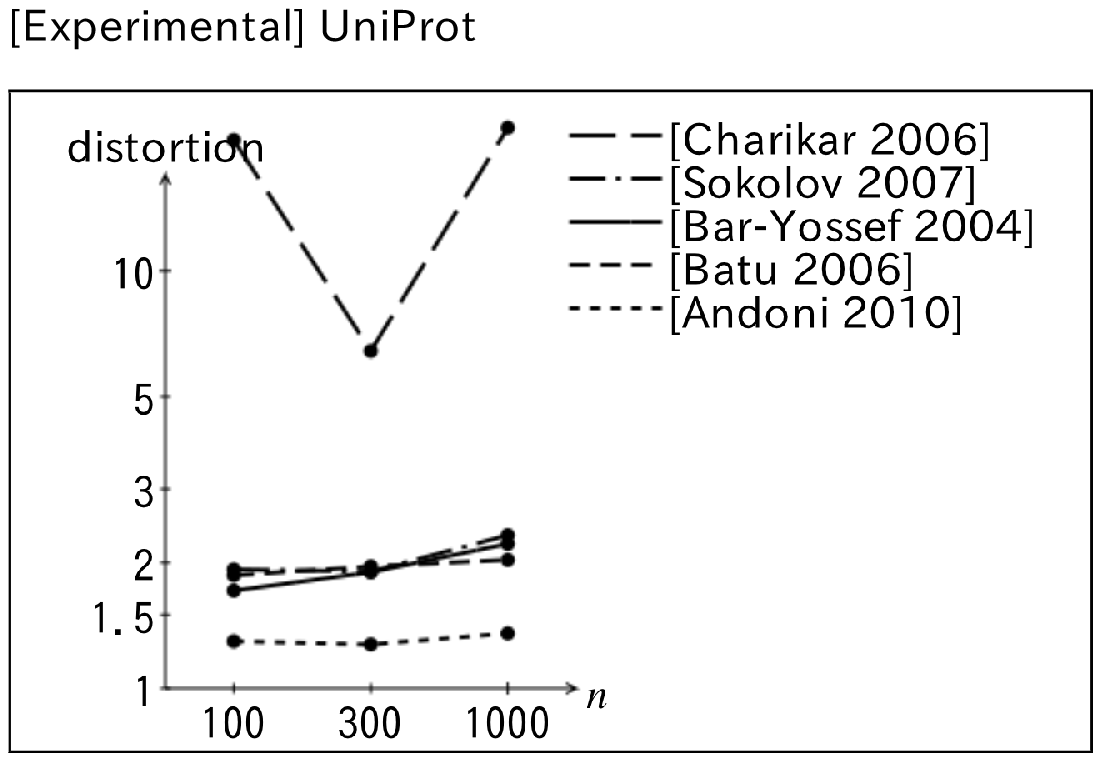}
\\
(e) DDBJ ($|\Sigma|\sim 4$, $e\sim n$) & (f) UniProt ($|\Sigma|\sim 20$, $e\sim n$)
\end{tabular}
}
\end{center}
\vspace{-1em}
\caption{Distortions of six data sets. Distortions larger than 30 are omitted from the charts.}
\label{fg:exp-datasets}
\end{figure}

We show the experimental results in Fig. \ref{fg:exp-algorithms}, Fig. \ref{fg:exp-datasets} and Table \ref{tb:best-approximation-exp}. From Fig. \ref{fg:exp-algorithms} we see that actual values of distortion are far less than their theoretical values, often 10 times or more (one scale mark in Fig. \ref{fg:exp-algorithms}). This is mainly because theoretical distortions are obtained in the worst case but real data are not the case.

We also see from Fig. \ref{fg:exp-algorithms} that the behavior (the outline of curves) obeys well the theoretical prediction, especially in [Batu 2006] and [Andoni 2010], whose asymptotic distortions are $O(1)$ and $O(\log n)$ under the condition of this experiment, respectively.

\begin{table}[tp]
\caption{Best algorithm according to the alphabet size $|\Sigma|$, the string length $n$ and the number of edits (an upper bound of the edit distance) $e$.}
\label{tb:best-approximation-exp}
\begin{center}
\includegraphics[width=12cm,clip]{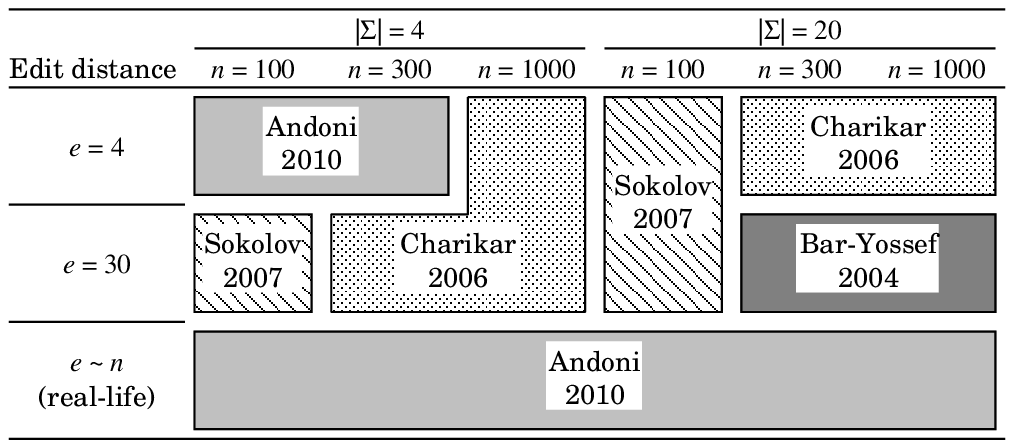}
\end{center}
\end{table}

Then we list the best algorithms depending on $|\Sigma|$, $n$ and $e$ in Table \ref{tb:best-approximation-exp} and the detailed comparison in Fig. \ref{fg:exp-datasets}. We assumed ``$e\sim n$'' in the two real-life data sets (DDBJ and UniProt) in Table \ref{tb:best-approximation-exp}, since they contain strings coming from many organic components and thus most string pairs have large (nearly $n$) edit distance.

We can see that [Andoni 2010], theoretically the second best, is almost always the best: it is the best for the two real-life data sets (DDBJ and UniProt) and nearly the best even for Random data set. On the other hand, theoretically the best algorithm [Batu 2006] did not yield the smallest distortion for any data set. Rather, as seen in Table \ref{tb:best-approximation-exp}, [Bar-Yossef 2004], [Charikar 2006] or [Sokolov 2007] becomes the best for Random data sets. Indeed, from Fig. \ref{fg:exp-datasets}, the conditions under which these algorithms achieved the smallest or near distortion are $|\Sigma| = 20$ for [Bar-Yossef 2004] and [Sokolov 2007], and $e = 4, 30$ for [Charikar 2006]. The possible explanation of their good achievements is as follows:
\begin{itemize}
\item {}[Bar-Yossef 2004] and [Sokolov 2007] showed better results for relatively large $|\Sigma|$. This is because they are $q$-gram-based algorithms. When $|\Sigma|$ is large, $q$-grams over $\Sigma$ appearing in a string become more distinct even if the value of $q$ is small. This means that the effect of appearance order\footnote{A counter example is $x = \text{``abcdefgh''}$ and $y = \text{``efghabcd''}$: the difference of appearance order makes the edit distance be large ($d_e(x, y) = 8 = n$) while 2-gram distance \cite{VectorRepSimilarString} is small (2).} disappears and thus $q$-gram distance becomes close to the edit distance.
\item {}[Charikar 2006] showed better results for $|\Sigma| = 20$ or ($|\Sigma| = 4$ and $e = 30$). This is because the distortion due to the alphabet expansion (Section \ref{ch:experiment-procedure}) can be small. When $|\Sigma|$ is large or $e$ is not so small compared with $n$, the expansion length $t$ to satisfy the Ulam condition can be small, especially in Random data set because uniform randomness works well.
\end{itemize}

We have analyzed only the distortion so far. However, there is a trade-off between the distortion and the computational cost. The computational costs of the six algorithms ranges from $O(n)$ ([Bar-Yossef 2004], [Charikar 2006] and [Sokolov 2007]) to $O(n^{1+\varepsilon})$ ([Batu 2006], [Andoni 2009] and [Andoni 2010]). In addition, in the latter three algorithms, we can control the trade-off by changing the value of $\varepsilon$. Since we carried out the experiment with $\varepsilon\sim 1$ (i.e. the least distortion at the expense of large time complexity $O(n^2)$ same as the edit distance), it might be better to take into account the time complexity for choosing an algorithm in practical problems.

\section{Conclusion}

We have compared six approximation algorithms of the edit distance in distortion, a measure of approximation accuracy, from the practical point of views: theoretical distortions without big-oh (asymptotic) notations, and experimental distortions in artificial and real-life data.

By the theoretical comparison, we have revealed the conditions on the string length $n$ for which these algorithms work best. The asymptotically best algorithm, [Batu 2006], was practically the best for $n\geq 300$, while [Bar-Yossef 2004] was the best for smaller $n$. In the experimental comparison, however, [Batu 2006] did not yield the best distortion for any data set, while [Andoni 2010] was the best or nearly best for most of real data sets, and [Bar-Yossef 2004], [Charikar 2006] and [Sokolov 2007] were the best or nearly best for large $|\Sigma|$. Since they are faster than [Batu 2006] and [Andoni 2010], it is worth changing the algorithm depending on the problems at hand.

The contribution of the paper is that this analysis revealed the ranges of $n$ where each approximation algorithm works better than the others with the absolute value of distortion, and that the experimental results revealed a large gap between theoretical and practical values of distortion in the algorithms.

In the future work, in addition to the discussion on the computational cost, we will narrow the gap between theoretical and experimental distortions by controling $d_e$ and $\theta$ in more detail (Section \ref{ch:compare-theoretical-distortions} and \ref{ch:exp-results}). We are also planning to apply them for real-life applications like biological sequence analyses, signal processing, or logging data analyses to confirm the accuracy and the computational time are practical enough.

\appendix
\section*{Appendix}
\section{Details of the distortion refinement without the big-oh notation} \label{ch:distortion-re-analyses}

Let $\log_b^* x$, called the {\it iterated logarithm} \cite{AlgoIntro-en}, be the minimum $i\geq 0$ such that $\displaystyle\underbrace{\log_b(\log_b(\dots \log_b x))}_{i\text{~`log's}} \leq 1$. If $x\leq 1$ then $\log_b^* x \defeq 0$. $\log_b^* x$ grows very slowly compared to $x$, e.g. $\lg^* x = 3$ if $x\in(4, 16]$ and $\lg^* x = 4$ if $x\in(16, 65536]$.

\subsection{[Batu 2006]} \label{ch:distortion-re-analyses-batu}

In Batu's algorithm \cite{ObliviousStrEmbed}, we first divide a string $x$ into blocks of length $c$ to $2c-1$ and compute the edit distance block-wise (i.e. treating a block as a character). As a result, the computational cost becomes $O((n/c)^2)$ after one division. The algorithm has two parameters $c\geq 2$, $j\geq 1$.\footnote{There is another parameter $\ell$, but we fixed $\ell = 1$ since it is enough for the single use of the distance (\cite{ObliviousStrEmbed}, pp. 799).} $j$ describes the number of the {\it alphabet reductions} (a string conversion process that only determines the boundaries of blocks). Note that we need to increase $c$ in accord with $n$ by $c = \omega(1)$ to assure $o(n^2)$-time computation. The authors of the paper take $c = (\lg\lg n)/\lg\lg\lg n$ (the end of Section 5 of \cite{ObliviousStrEmbed}). In Section \ref{ch:theoretical-distortions} we took $c = \max\{\lg\lg c/(\lg\lg\lg c), 2\}$ instead. In Section \ref{ch:experimental-distortions} we fixed $c = 2$ for the theoretical distortion since we took only $c = 2, 4$ for the experiment.

The distortion $K$ is given by
\begin{eqnarray}
K&=& (2c-1)\cdot O((3c^2\log c)^c/c + \log^* kc) \label{ex:batu2006-distortion-ell1} \\
 & & (\text{Theorem 4.1 in \cite{ObliviousStrEmbed}, pp. 797}) \nonumber \\
 &=& (2c-1)\cdot [4c(\log^* kc + O(1)) + O((3c^2\log c)^c)]/c \nonumber \\
 & & (\text{Lemma 4.5 in \cite{ObliviousStrEmbed}, pp. 797}) \nonumber \\
 &=& (2c-1)\cdot [4cj + O((3c^2\log c)^c)]/c. \label{ex:batu2006-eval-ell1} \\
 & & (\text{Lemma 4.5 in \cite{ObliviousStrEmbed}, pp. 797}) \nonumber
\end{eqnarray}
where $k = \lceil\lg|\Sigma|\rceil$ is the number of bits to describe a character. The remained big-oh notation $O((3c^2\log c)^c)$ is evaluated as follows: $O((3c^2\log c)^c)$ is obtained from $2^{k_j}$ where $k_i = (c - 1)\cdot(\lceil\lg((2c-3)k_{i-1})\rceil + 2)$, $k_0 = k$ (pp. 796 in \cite{ObliviousStrEmbed}). 

\subsubsection{The case of $j=1$}

If $j=1$, used in Section \ref{ch:experimental-distortions}, then $k_1 = (c - 1)\cdot(\lceil\lg((2c-3)k_{i-1})\rceil + 2)\leq (c - 1)\cdot(\lg((2c-3)k) + 3)$ and thus the distortion becomes
\begin{eqnarray}
K\leq (2c-1)\cdot [4c + \{8(2c-3)k\}^{c-1}]/c. \label{ex:batu2006-distortion-once}
\end{eqnarray}

\subsubsection{The case $j$ is large enough}

Then we consider the case $j$ is large enough for the small distortion. In this case $k_j$ becomes the fixed point of $k_i = (c - 1)\cdot(\lceil\lg((2c-3)k_{i-1})\rceil + 2)$. We can easily confirm that $k_j\leq 4(c - 1)^2$ since it is larger than $(c - 1)\cdot(\lceil\lg((2c-3)k)\rceil + 2)$ for any $c\geq 2$.\footnote{We found an upper bound $\hat{k_j} = 4(c - 1)^2$ as follows: since $k$ is asymptotically larger than $(c - 1)\cdot(\lceil\lg((2c-3)k)\rceil + 2)$ in $k$, $\hat{k_j}$ must satisfy $\hat{k_j}\geq (c - 1)\cdot(\lceil\lg((2c-3)\hat{k_j})\rceil + 2)$. As a result, $k_j = \omega(c)$ is required. Thus we first take $k = \gamma(c - 1)^2$ and then supplied the constant $\gamma$ to satisfy the inequality.} In addition, $j$ is large enough with $\lg((2c-3)k) + 1$ if $k\geq k_j$ since the number of binary digits of $k_i$ in the recurrence is reduced by at least one except for the final recurrence. As a result, from the expression (\ref{ex:batu2006-eval-ell1}), an upper bound of the distortion becomes
\begin{eqnarray*}
K &=& (2c-1)\cdot [4cj + O((3c^2\log c)^c)]/c \\
  &\leq& 4(2c-1)\left(\lg((2c-3)k) + 1 + \frac{(c - 1)^2}{c}\right).
\end{eqnarray*}
\subsection{[Charikar 2006]} \label{ch:reanalysis-charikar}

The distortion of Charikar's method \cite{EmbedUlamMetric} is evaluated as $O(\log n)$ for Ulam metric. First we show its value without big-oh notation. The approximation function $\|f(P)-f(Q)\|$, where $P$ and $Q$ are strings satisfying the Ulam condition, is evaluated as follows in \cite{EmbedUlamMetric}:
\begin{eqnarray*}
\|f(P)-f(Q)\|&\leq& 3(1+\ln n)\leq 3(1+\ln n)\frac{d_e(P, Q)}{\theta} \\&&\text{(if $P\neq Q$; in Lemma 2.2, pp.211 in \cite{EmbedUlamMetric})}\\
\|f(P)-f(Q)\|&\geq& d_e(P, Q)/8 \\&&\text{(in Lemma 2.3, pp.212 in \cite{EmbedUlamMetric})}
\end{eqnarray*}
Thus we get $d_e(P, Q)/8\leq\|f(P)-f(Q)\|\leq 3(1+\ln n)\frac{d_e(P, Q)}{\max\{1,\theta\}}$, where $\theta$ is replaced with $\max\{1,\theta\}$ since the expression above does not consider the case $d_e(P, Q) = 0$. This concludes the distortion of $\|f(P)-f(Q)\|$ for the Ulam metric is $\frac{24(1+\ln n)}{\max\{1,\theta\}}$.

In addition, in the manner in Section \ref{ch:experiment-procedure}, the distortion for any strings is $\frac{24(1+\ln n)}{\max\{1,\theta\}}\cdot 2n = \frac{48n(1+\ln n)}{\max\{1,\theta\}}$ since $t$ is at most $n$.

\subsection{[Andoni 2009]}

The distortion for [Andoni 2009] \cite{L1NonEmbeddability} is concluded as $O(1)$ for the Ulam metric. We have removed the big-oh notation as follows: The approximation function $d_{\mathrm{NEG},\infty,1}(\phi(P),\phi(Q))$, where $P$ and $Q$ are strings satisfying the Ulam condition, is evaluated as follows in \cite{L1NonEmbeddability}:

\begin{eqnarray*}
d_{\mathrm{NEG},\infty,1}(\phi(P),\phi(Q)) &\geq& \underline{d_e}(P,Q)/50 \\
	&&\text{(Proof of Theorem 1.1, pp.870)}\\
d_{\mathrm{NEG},\infty,1}(\phi(P),\phi(Q)) &\leq& 17\underline{d_e}(P,Q) \\
	&&\text{(Proof of Theorem 1.1, pp.871)}\\
\underline{d_e}(P,Q)\leq d_e(P,Q) &\leq& 2\underline{d_e}(P,Q) \\
	&&\text{(Section 1.5, pp.868)}
\end{eqnarray*}

As a result, the distortion for Ulam metric is calculated as $50\cdot 17\cdot 2=1700$. In addition, in the manner in Section \ref{ch:experiment-procedure}, the distortion for any strings is $1700\cdot 2n = 3400n$ since $t$ is at most $n$.

\section{Details of the distortion calculation from inequalities} \label{ch:distortion-from-ineq}

\subsection{[Bar-Yossef 2004]}

The upper and the lower bounds of \cite{ApproxEditDistEfficiently}=[Bar-Yossef 2004] are given by
\begin{eqnarray*}
\begin{cases}
d_e\leq k \Rightarrow \tilde{d_e}\leq 4kq, \\
d_e\geq 13(kn)^{\frac{2}{3}} \Rightarrow \tilde{d_e}\geq 8kq. ~~~~\text{(with $q=n^{2/3}/(2k^{1/3})$)}
\end{cases}
\end{eqnarray*}
As a result we obtain
\begin{eqnarray*}
\frac{4}{13}d_e\leq\tilde{d_e}\leq 2(d_e n)^{2/3}.
\end{eqnarray*}
As shown in Section \ref{ch:calc-distortion-asympt}, since $u(d_e)/d_e = 2(n^2/d_e)^{1/3}$ and $l(d_e)/d_e = 4/13$ are monotonically decreasing and increasing, respectively, the distortion for $d_e\geq\theta$ is $K_{\theta} = u(\theta)/l(\theta) = 2(\theta n)^{2/3}/(\frac{4}{13}\theta) = (13n^{2/3})/(2\theta^{1/3})$.

\subsection{[Sokolov 2007]}

The upper and the lower bounds of \cite{VectorRepSimilarString}=[Sokolov 2007] are given by
\begin{eqnarray}
d_e(x, y)\leq k &\Rightarrow& \tilde{d_e}(x, y)\leq (2k(n+2))/n, \nonumber\\
d_e(x, y) > k &\Rightarrow& \tilde{d_e}(x, y)\geq 2(k - 4)/n. \label{ex:sokolov-lbound}
\end{eqnarray}

Note that the distortion should be treated as $+\infty$ if $\theta \leq 5$ since $\tilde{d_e}(x, y)$ can be zero if $d_e(x, y)$ is less than 5, that is, $k$ is less than 4, from (\ref{ex:sokolov-lbound}). Otherwise we obtain
\begin{eqnarray*}
2(d_e - 5)/n\leq\tilde{d_e}\leq (2d_e(n+2))/n.
\end{eqnarray*}
As shown in Section \ref{ch:calc-distortion-asympt}, since $u(d_e)/d_e = (2d_e(n+2))/(n d_e)$ and $l(d_e)/d_e = 2(d_e - 5)/(n d_e)$ are monotonically decreasing and increasing, respectively, the distortion for $d_e\geq\theta$ is $K_{\theta} = u(\theta)/l(\theta) = [(2\theta(n+2))/n]/[2(\theta - 5)/n] = (\theta(n+2))/(\theta - 5)$.



\bibliographystyle{unsrt}
\bibliography{string}







\end{document}